\documentclass{article}

\PassOptionsToPackage{numbers, compress}{natbib}
\usepackage{graphicx}

\usepackage[preprint]{neurips_2024}

\usepackage[utf8]{inputenc} 
\usepackage[T1]{fontenc}    
\usepackage{hyperref}       
\usepackage{url}            
\usepackage{booktabs}       
\usepackage{amsfonts}       
\usepackage{nicefrac}       
\usepackage{microtype}      
\usepackage{xcolor}         
\usepackage{graphicx}
\usepackage{multirow}
\usepackage{xcolor}
\usepackage{subfigure}
\usepackage{authblk}
\title{YingSound: Video-Guided Sound Effects Generation with Multi-modal Chain-of-Thought Controls}
\author[1]{Zihao Chen}
\author[1]{Haomin Zhang}
\author[1]{Xinhan Di}
\author[1,3]{Haoyu Wang}
\author[1,3]{Sizhe Shan}
\author[1]{Junjie Zheng}
\author[1]{Yunming Liang}
\author[1,4]{Yihan Fan}
\author[1,2]{Xinfa Zhu}
\author[1,2]{Wenjie Tian}
\author[1]{Yihua Wang}
\author[1]{Chaofan Ding}
\author[2]{Lei Xie}
\affil[1]{AI Lab, Giant Network}
\affil[2]{ASLP Lab, Northwestern Polytechnical University}
\affil[3]{Zhejiang University}
\affil[4]{East China University of Science and Technology}

\begin{document}
\maketitle
\date{}

\begin{abstract}
Generating sound effects for product-level videos, where only a small amount of labeled data is available for diverse scenes, requires the production of high-quality sounds in \textit{few-shot} settings. To tackle the challenge of limited labeled data in real-world scenes, we introduce \textit{YingSound}, a foundation model designed for video-guided sound generation that supports high-quality audio generation in few-shot settings. Specifically, YingSound consists of two major modules. The first module uses a conditional flow matching transformer to achieve effective semantic alignment in sound generation across audio and visual modalities. This module aims to build a learnable audio-visual aggregator (AVA) that integrates high-resolution visual features with corresponding audio features at multiple stages. The second module is developed with a proposed multi-modal visual-audio chain-of-thought (CoT) approach to generate finer sound effects in few-shot settings. Finally, an industry-standard video-to-audio (V2A) dataset that encompasses various real-world scenarios is presented. We show that YingSound effectively generates high-quality synchronized sounds across diverse conditional inputs through automated evaluations and human studies. Project Page: \url{https://giantailab.github.io/yingsound/}

\end{abstract}

\section{Introduction} 
In recent years, audio, speech, and music generation development has advanced unprecedentedly. Speech generation has seen significant progress, particularly in two distinct areas: traditional text-to-speech (TTS) systems~\cite{ren2020fastspeech,wang2017tacotron,kim2021conditional,tan2024naturalspeech,wang2018style,lee2019robust,shen2023naturalspeech,sotelo2017char2wav,ye2024flashspeech} and large-scale speech models~\cite{zhang2023speechgpt,wang2023neural,fang2024llama,xie2024mini,defossez2024moshi,du2023lauragpt,fathullah2024audiochatllama,fu2024vita,chu2023qwenaudioadvancinguniversalaudio}. Traditional TTS systems are typically trained on limited datasets recorded in studios, lacking high-quality zero-shot capabilities. In contrast, large-scale speech models, trained on extensive datasets, exhibit strong zero-shot performance and deliver superior prosody and naturalness. These advancements collectively contribute to a broad range of applications. Similarly, music generation has emerged as a vibrant field, encompassing both symbolic music generation based on scores or MIDI~\cite{dong2018musegan,dong2023multitrack,mittal2021symbolic,wu2024generating,lv2023getmusic,yang2017midinet} and waveform generation conditioned on text or visual prompt~\cite{agostinelli2023musiclm,copet2024simple,di2021video, dhariwal2020jukebox,lam2024efficient}. These approaches have shown remarkable potential. However, the development of multi-modal based audio generation remains in its early stages.

Audio generation has immense potential for applications like Foley sound production, gaming, virtual reality (VR), and background audio creation. Advances in generative models have made it possible to synthesize audio from multi-modal (video \& audio \& text) inputs. For instance, latent diffusion models enable audio synthesis conditioned on text by controlling the diffusion process, while flow matching~\cite{guan2024lafma} and consistency models~\cite{liu2024audiolcm} offer alternative approaches to audio modeling. Furthermore, treating audio generation as a language modeling task~\cite{borsos2023audiolm} has opened new avenues for research and application.

Synthetic audio can become more vivid and contextually rich when guided by images~\cite{sheffer2023hear} and videos~\cite{SpecVQGAN_Iashin_2021}. Moreover, the temporal co-occurrence between video and audio provides a natural alignment for audio synthesis. This makes video a powerful guide to generating audio, enabling applications such as AI-generated video dubbing and Foley sound production for films. These capabilities underline the promising and broad application potential of V2A generation.

In filmmaking, Foley reproduces everyday sound effects added to films, videos, and other media in post-production to enhance audio quality. Traditional Foley requires a significant amount of human effort and time. Recently, with the rapid development of video generation~\cite{polyak2024movie, ho2022imagen, singer2022make, hong2022cogvideo, yang2024cogvideox, kong2024hunyuanvideo}, designers can quickly generate video clips according to their imagination. However, the generated video clips are often silent without background sound. To address this issue, researchers have started using V2A techniques to generate audio associated with video. Moreover, V2A can be applied to create sound effects for short videos and professional film and television productions. 

Initially, text-to-audio (T2A) technology~\cite{yang2023uniaudio,yang2024uniaudio15largelanguage, audioldm2-2024taslp, xue2024auffusion,borsos2023audiolmlanguagemodelingapproach,liu2024m2ugenmultimodalmusicunderstanding,huang2023make,xie2024picoaudio} is commonly used to generate corresponding sound effects, which are then manually aligned with video. This approach still requires a lot of 
 effort with prompt engineering and manual alignment (both semantically and temporally). In addition, other key factors include sound quality and fine-grained generation. 

This paper presents YingSound, a video-guided sound effects generation approach with multi-modal CoT controls. It consists of a transformer-based flow matching module with a learnable AVA connector, a multi-modal visual-audio CoT module, and a corresponding training strategy enabling audio generation towards few-shot settings. Our key contributions are as follows:

\begin{itemize}
\item We design a learnable audio-visual aggregator (AVA), a dynamic module that integrates high-resolution visual features with corresponding audio features at multiple stages within the flow matching transformers.
\item We propose a multi-modal visual-audio CoT module that generates high-quality audio in few-shot settings for industrial scenes optimized through reinforcement learning.
\item We develop an industry-standard video-to-auido (V2A) dataset encompassing various real-world scenarios, including movies, games, and commercials.
\end{itemize}

\section{Related Work}
\textbf{Audio-visual learning.} Audio-visual learning has been a central focus in multi-modal representation research, with numerous studies exploring the relationship between audio and visual data. Some studies explore the semantic connections between sight and sound, focusing on learning shared audio-visual associations~\cite{arandjelovic2017look, morgado2021audio, girdhar2023imagebind}, audio-visual sound localization~\cite{arandjelovic2018objects, hu2022mix}, and audio-visual segmentation~\cite{zhou2024audio}. Other research emphasizes the temporal alignment of audio and video events~\cite{chen2021audio, feng2023self}. 
Similarly, we develop a learnable AVA, a module that integrates high-resolution visual features for V2A generation, which achieves semantic and temporal alignment between audio and vision progressively across multiple stages. 

\textbf{Video-to-audio Generation.} The current state-of-the-art V2A models can be divided into two main approaches: one involves training end-to-end models directly on large-scale datasets, while the other builds upon the base capabilities of T2A models, enhancing them with temporal and video-semantic alignment.
For end-to-end models, two common strategies have emerged. The first employs diffusion models to generate Mel spectrograms guided by visual features directly~\cite{comunità2023syncfusionmultimodalonsetsynchronizedvideotoaudio,xu2024videotoaudiogenerationhiddenalignment,wang2024tiva,cheng2024lovalongformvideotoaudiogeneration,ren2024stav2avideotoaudiogenerationsemantic,luo2023difffoley,wang2024frieren}. The second approach uses transformer architectures to autoregressively predict audio tokens, which are then fed into a decoder to generate audio~\cite{viertola2024temporally,iashin2021tamingvisuallyguidedsound,sheffer2023iheartruecolors,du2023conditionalgenerationaudiovideo,mei2023foleygenvisuallyguidedaudiogeneration}.  Notable examples of these approaches include Frieren~\cite{wang2024frieren}, which employs rectified flow matching with a non-autoregressive vector field estimator, integrating feedforward transformers and cross-modal feature fusion to achieve robust temporal alignment. Movie Gen Audio~\cite{polyak2024movie} explores generating audio conditioned on video and text. For the two-stage methods, current research~\cite{xie2024sonicvisionlm,pascual2024masked,zhang2024foleycrafter,jeong2024read,yang2024draw,xing2024seeinghearingopendomainvisualaudio,wang2023v2amapperlightweightsolutionvisiontoaudio,yi2024efficientvideoaudiomapper} leverages the capabilities of T2A models, incorporating temporal or semantic alignment to achieve high-quality V2A. For example, SonicVisionLM~\cite{xie2024sonicvisionlm} aims to generate diverse sound effects by utilizing vision-language models (VLMs). MaskVAT~\cite{pascual2024masked} proposes a full-band, high-quality general audio codec with a sequence-to-sequence masked generative model. FoleyCrafter~\cite{zhang2024foleycrafter} and Rewas~\cite{jeong2024read} leverage a pre-trained text-to-audio model to ensure high-quality audio generation. FoleyCrafter and Rewas adopt time and energy conditions, respectively, to guide sound effects generation. Draw-an-audio~\cite{yang2024draw} proposes the mask-attention module (MAM) that focuses on regions of interest using masked video instructions. At the same time, the time-loudness module (TLM) aligns sound synthesis with the video in terms of loudness and timing. However, the above methods lack exploration of audio generation in few-shot settings. Therefore, we design a multi-modal visual-audio CoT module to facilitate audio generation in such settings.

\section{Approach}

\subsection{Overview}
We present a V2A model, YingSound, designed for industrial-level audio generation from videos and address challenges such as long-form audio generation with limited annotations. YingSound adopts a few-shot learning approach and involves three key aspects. Firstly, a comprehensive data collection and annotation pipeline is established to build a high-quality task-specific dataset at the industrial-level. Secondly, a learnable AVA is developed, generating audio from video inputs. The AVA is a two-tower transformer network that leverages the capability of a pre-trained T2A model and progressively builds audio-visual alignment across multiple training stages. Thirdly, a CoT module is introduced to refine the audio generated by the AVA in few-shot settings, which produces high-quality, contextually appropriate audio through an expert module with multi-modal reasoning.
\begin{figure}[htbp]
\centering
\includegraphics[width=1\textwidth]{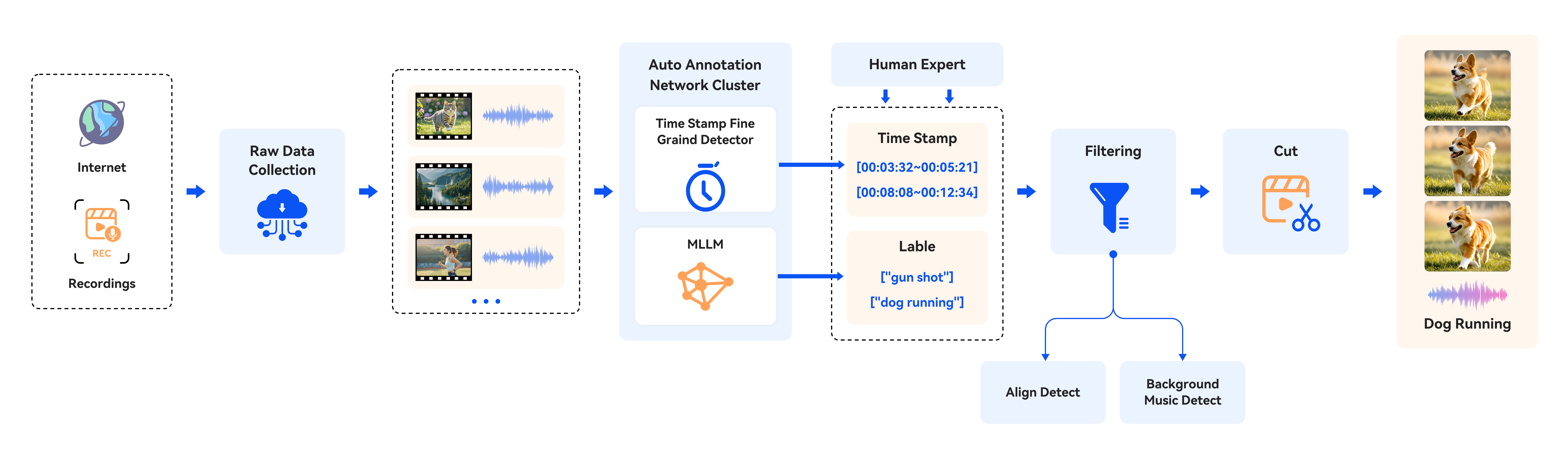}
\caption{ The data collection and processing pipeline with human-in-the-loop.}
\label{fig:1}
\end{figure}
\subsection{Data-Pipeline}
As illustrated in Figure \ref{fig:1}, we design a comprehensive data processing pipeline that includes collection, annotation, filtering, and cutting. It is essential to integrate human expertise with AI, as the complexity of the entire process and the variation across different video types, relying solely on automation, are insufficient. \\
\textbf{Data Collection and Annotation.} We start by classifying video types and then download or record content from the internet based on these categories to improve diversity. Each video comes with its corresponding audio. To ensure data quality, we resample the audio and convert it to mono. We develop an automated annotation network cluster, comprising a fine-grained timestamp detector and multi-modal large language models (MLLMs) for timestamp extraction and text labeling. We utilize advanced video-large-language models (Video-LLMs) for complex video understanding, accurately annotating multiple sound events. Subsequently, the detected sound events from the previous step are fed into a fine-grained audio timestamp prediction model, producing precise timestamp for seamless integration into downstream tasks. In addition, certain tasks require manual annotation to ensure accuracy and handle complex cases effectively. \\
\textbf{Filtering and Cutting.} Data with low alignment quality, including temporal alignment and video-semantic alignment, or strong background music, is discarded to ensure the overall quality of the dataset. We use the AV-Align~\cite{yariv2024diverse} metric to filter out data with low temporal alignment scores, ensuring that only high-quality audio-video pairs are retained for further processing. Next, we employ CLIP and CLAP~\cite{elizalde2023clap} techniques to filter out semantically misaligned data. To eliminate pure background music and speech-involved segments, we utilize background music detection techniques and automatic speech recognition (ASR)~\cite{gao2023funasr, yao2021wenet,radford2023robust} to filter out audio with non-relevant content. Finally, after multiple rounds of filtering, we cut the data based on precise timestamp and label annotations, resulting in a high-quality audio-video-text dataset ready for subsequent training. \\
\begin{figure}[h]
\centering
\includegraphics[width=1\textwidth]{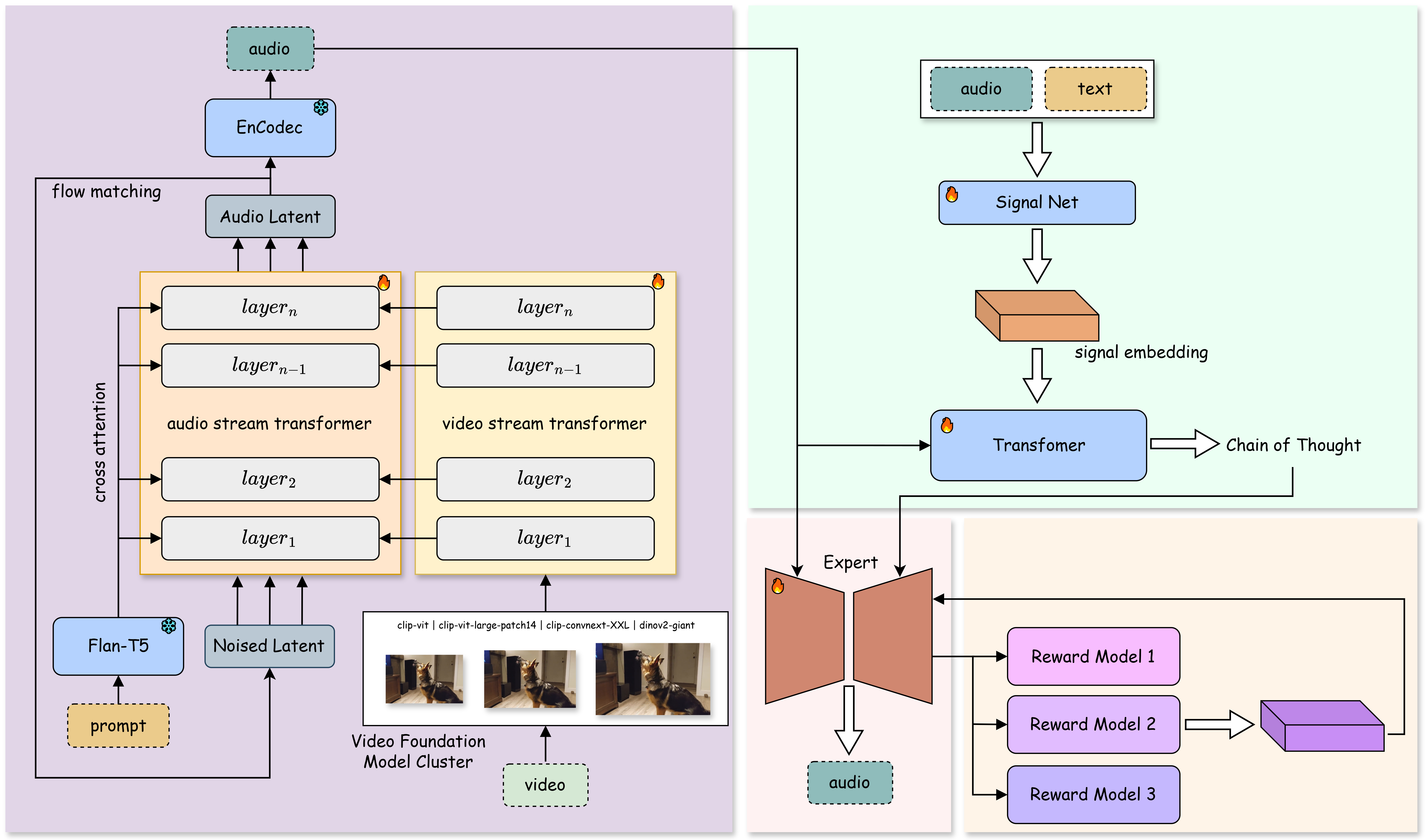}
\caption{ \textbf{The overview of the YingSound.} It comprises two key components: Conditional Flow Matching with Transformers and a Multi-modal Chain-of-Thought Based Audio Generation.}
\label{fig:2}
\end{figure}

\subsection{Conditional Flow Matching with Transformers}
We perform audio generation based on flow matching~\cite{yang2024consistency} and a scalable diffusion transformer (DiT) architecture~\cite{peebles2023scalable}, which have been proven to be both high-quality and efficient in image and audio generation. Our model can be conditioned on both text prompts and videos. For text conditions, we adopt an instruction-tuned LLM FLAN-T5~\cite{chung2024scaling} as a text encoder and apply cross attention in every audio DiT layer similar to Tango~\cite{ghosal2023tango} and some other diffusion-based models. For video conditions, we construct a Video Foundation Model Cluster inspired by~\cite{shi2024eagle}. We use different input resolutions and different visual encoders pre-trained, including clip-vit, clip-vit-large-patch14, clip-convnext-XXL~\cite{schuhmann2022laionb}, dinov2-giant~\cite{oquab2023dinov2} to get frame-level features. A linear layer then projects the clip features to a suitable dimension for the video DiT.

We encode the raw waveform with Encodec~\cite{defossez2022high}, which is trained in 24 kHz monophonic audio in diverse domains, including speech, music, and general audio. In particular, we use the features extracted before the residual quantization layer~\cite{kreuk2022audiogen,vyas2023audiobox,copet2024simple}.

\paragraph{Audio-Visual Aggregator Module.}
Video frames contain rich information about the audio content of each frame. We use an audio video mapping Module (AVMM) between every DiT layer to perform cross-modality mapping~\cite{burtsev2021multi}. Given an output of audio/video DiT layer i \(y_a^i,y_v^i\), the input of the next layer \(x_a^{i+1},x_v^{i+1}\) is computed as:
$$x_a^{i+1}=y_a^i+Linear_a(concat(y_a^i, y_v^i))$$
$$x_v^{i+1}=y_v^i+Linear_v(concat(y_a^i, y_v^i))$$

\paragraph{Multi-Stage Training for the T2A/V2A Generation.}
We train our model from T2A to V2A step by step to make the training process more stable.
\begin{itemize}
\item First, the model is trained only using the T2A data, which means that the video condition module is omitted, including the video DiT. In this stage, we obtain a sufficiently good T2A model.

\item Videos are then added to train the video condition module, and video descriptions are retained during training. The ratio of T2A and T\&V2A training data is kept at 1:1 during this stage.

\item Finally, we randomly drop some text prompts for videos so that the model can generate audio purely from video. The T2A, T\&V2A, and V2A training data ratio is set to 1:1:2 during this stage.
\end{itemize}
So, the probability of keeping the text conditions for each stage is 100\%/100\%/50\%. And the probability of keeping the video conditions is 0\%/50\%/75\%. Finally, we get a model capable of generating audio based on texts, videos, or both.

\subsection{Multi-modal Chain-of-Thought Based Audio Generation}
To enable audio generation from guided video at an industrial-level, we address the challenge of the high cost associated with fully supervised training samples. We propose a multi-modal CoT generation module~\cite{gao2024cantor,guo2024iw,deng2024r,hu2024visual} tailored for few-shot settings. This module comprises two key components. The first component is the multi-modal CoT module, which generates multi-modal CoT based on coarse-level audio outputs. The second component focuses on finer-level audio generation, leveraging the outputs from the coarse audio and the corresponding CoT with the application of reward models 
~\cite{rocamonde2023vision,fu2024furl,wang2024rovrm,wang2024rl,chen2021learning,baumli2023vision,li2024auto}. 

\paragraph{Multi-modal Chain-of-Thought Module.}
The multi-modal CoT module~\cite{mondal2024kam,zheng2023ddcot,suris2023vipergpt} consists of two components. The first component is a signal module, which receives multi-modal inputs and outputs signal embeddings. The second component is the generation module, which inputs coarse-level audio and signal embeddings and produces the corresponding CoT for finer-level audio. Both components are built with multiple layers of transformers.  

\paragraph{Finer-level Audio Generation.}
The audio generation module at the finer-level consists of two components. The first component is a reward module, composed of a set of reward models to generate different rewards based on the audio. The second component is an expert module that generates finer-level audio that aligns with the rewards, corresponding CoT, and coarse-level audio.

\paragraph{Multi-Stage Training for Adaptive Generation.}
To enable adaptive generation in few-shot settings, the training process follows these steps:
\begin{itemize}
\item First, we train the two components in a few-shot supervised setting as an initial training step, using supervised samples. 

\item Second, we proceed with the training stage using generated CoT in an iterative preference learning process to minimize the generation of incorrect CoT.   

\item Third, we transition to joint training in an intrinsic self-correction setting to improve audio quality iteratively.    
\end{itemize}
\section{Experiments}

\begin{table}
\caption{Dataset Details.}
\centering
\label{tab:1}
\begin{tabular}{cccc}
\toprule
Dataset & Traning Stages & Modality & Clip Numbers \\
\midrule
AudioCaps~\cite{kim2019audiocaps} & 1/2/3 & T/A & 49k \\
WavCaps~\cite{mei2024wavcaps} & 1/2/3 & T/A & 402k \\
TangoPromptBank~\cite{ghosal2023tango} & 1/2/3 & T/A & 37k \\
MusicCaps~\cite{agostinelli2023musiclm} & 1/2/3 & T/A & 5k \\
AF-AudioSet~\cite{gemmeke2017audio} & 1/2/3 & T/A & 695k \\
VGGSound~\cite{chen2020vggsound} & 2/3 & T/V/A & 173k \\
\bottomrule
\end{tabular}
\end{table}

\begin{table}
\caption{Objective results of VGGSound-Test regarding audio quality, semantic alignment, and temporal alignment. w. text denotes audio generation with text as a guiding condition, and w/o text denotes audio generation without text input, using only the video content.}
\centering
\label{tab:2}
\begin{tabular}{ccccccc}
\toprule
Method&FAD$\downarrow$&FD $\downarrow$&KL-sigmoid $\downarrow$&IS $\uparrow$ &CLIP $\uparrow$ & AV $\uparrow$\\
\midrule
Diff-Foley~\cite{luo2023difffoley}& 6.05 & 23.38 & 7.86  & 10.95 & 9.40  & 0.21\\
FoleyCrafter w/o text~\cite{zhang2024foleycrafter}& 2.38 & 26.70 & 6.24 & 9.66 & 15.57 & {\bfseries0.25}\\
FoleyCrafter w. text~\cite{zhang2024foleycrafter}& 2.59 & 20.88 & 5.38 & 13.60 & 14.80 & 0.24\\
V2A-Mapper~\cite{wang2024v2a}& 0.82 & 13.47 & 6.57 & 10.53 & 15.33 & 0.14 \\
Frieren~\cite{wang2024frieren}& 1.36 & 12.48 & 6.58 & 12.34 & 11.57 & 0.21 \\
YingSound w/o text (ours) \ &0.80  &  8.66    & 5.12 & 12.08 & 16.14 & {\bfseries0.25}\\
YingSound w. text (ours) \ & {\bfseries0.78} & {\bfseries6.28} &{\bfseries 3.97 }   & {\bfseries  14.02}  & {\bfseries 16.86}& {\bfseries0.25}\\
\bottomrule
\end{tabular}
\end{table}

\subsection{Experiment Settings}
\textbf{Dataset.} We use about 1.2M text-audio pairs, including AudioCaps~\cite{kim2019audiocaps}, WavCaps~\cite{mei2024wavcaps}, TangoPromptBank~\cite{ghosal2023tango}, MusicCaps~\cite{agostinelli2023musiclm}, and AF-AudioSet~\cite{gemmeke2017audio}, about 3.3k hours, to span the entire three-stage training process. We only use the VGGSound~\cite{chen2020vggsound} dataset as video-related input in the second and third training stages. Datasets are detailed in Table \ref{tab:1}.

\textbf{Implementation details.} We train the first stage model using the entire T2A dataset for a total of 250k steps. In the second stage, we utilize the complete VGGSound dataset along with its corresponding T2A data, where both text and audio are used for training. This phase involves a total of 50k training steps. In the third stage, we use a combination of T2A, T\&V2A, and V2A datasets in a ratio of 1:1:2, with a total of 230k training steps. The training is carried out with a batch size of 128 on 8 Nvidia A800 GPUs. We use the Adam optimizer with a learning rate of 3e-5. We also clip the gradient norm to 0.2 for training stability. During inference, we use a sway sampling strategy~\cite{chen2024f5} with NFE $= 64$ to improve our model's performance.

\textbf{Evaluation Metrics.} For evaluation, we employ several evaluation metrics to assess semantic alignment, temporal alignment, and audio quality, namely Inception Score (IS)~\cite{salimans2016improved}, CLIP score (CLIP), Fr\'{e}chet Distance (FD)~\cite{heusel2017gans}, Fr\'{e}chet Audio Distance (FAD), AV-align (AV) and KL Divergence-sigmoid (KL-sigmoid)~\cite{iashin2021taming} by calculating all data in the VGGSound-test set. The CLIP measures the similarity between the input video and the generated audio embeddings within the same representation space. To achieve this, we utilize Wav2CLIP~\cite{wu2022wav2clip} as the audio encoder and CLIP as the video encoder, following the approach adopted in prior studies~\cite{sheffer2023hear}. FD assesses the distribution similarity to evaluate the fidelity of the generated audio, using PANNs~\cite{kong2020panns} as the feature extractor. FAD evaluates the fidelity of the generated audio by assessing the similarity of its distribution by using VGGish~\cite{koh2021comparison} as the feature extractor. KL-sigmoid computes the average KL-divergence across all classes in the test set to assess the similarity at the paired sample level. IS assesses sample quality and diversity between ground truth and generated audio. AV is based on the detection and comparison of energy peaks in both modalities to assess the temporal alignment of the input audio and generated video. 

\textbf{Baselines.} We conduct comprehensive evaluations of YingSound by comparing it with state-of-the-art approaches, namely Diff-Foley, FoleyCrafter, V2A-Mapper~\cite{wang2024v2a}, and Frieren. Diff-Foley adopts contrastive audio-visual pre-training (CAVP)~\cite{jung2022cavp} to learn more temporally and semantically aligned features, then trains an LDM with CAVP-aligned visual features in the spectrogram latent space. FoleyCrafter adopts a novel framework that leverages a pre-trained text-to-audio model to ensure high-quality audio generation. FoleyCrafter comprises two key components: the semantic adapter for semantic alignment and the temporal controller for precise audio-video synchronization. We evaluate both with and without (FoleyCrafter w. or w/o text) text as a condition. Like FoleyCrafter, our model also supports two types of sound effects generation results: one with text input, where text serves as the condition for generating audio based on video content, and video-based audio generation (YingSound w. or w/o text). V2A-Mapper, evaluated on only 7,590 WAV clips, introduces a module to translate visual inputs for high-fidelity sound generation, achieving superior performance with fewer parameters compared to existing approaches. Frieren uses rectified flow matching to synthesize high-quality, temporally aligned audio from video. It features a non-autoregressive vector field estimator and channel-level cross-modal feature fusion, enabling efficient audio generation with superior synchronization and quality.

\begin{figure}[t]
	\centering
	\vspace{-0.15in}
	\begin{minipage}{1\linewidth}
		\subfigure{
			\includegraphics[width=0.49\linewidth,height=1.2in]{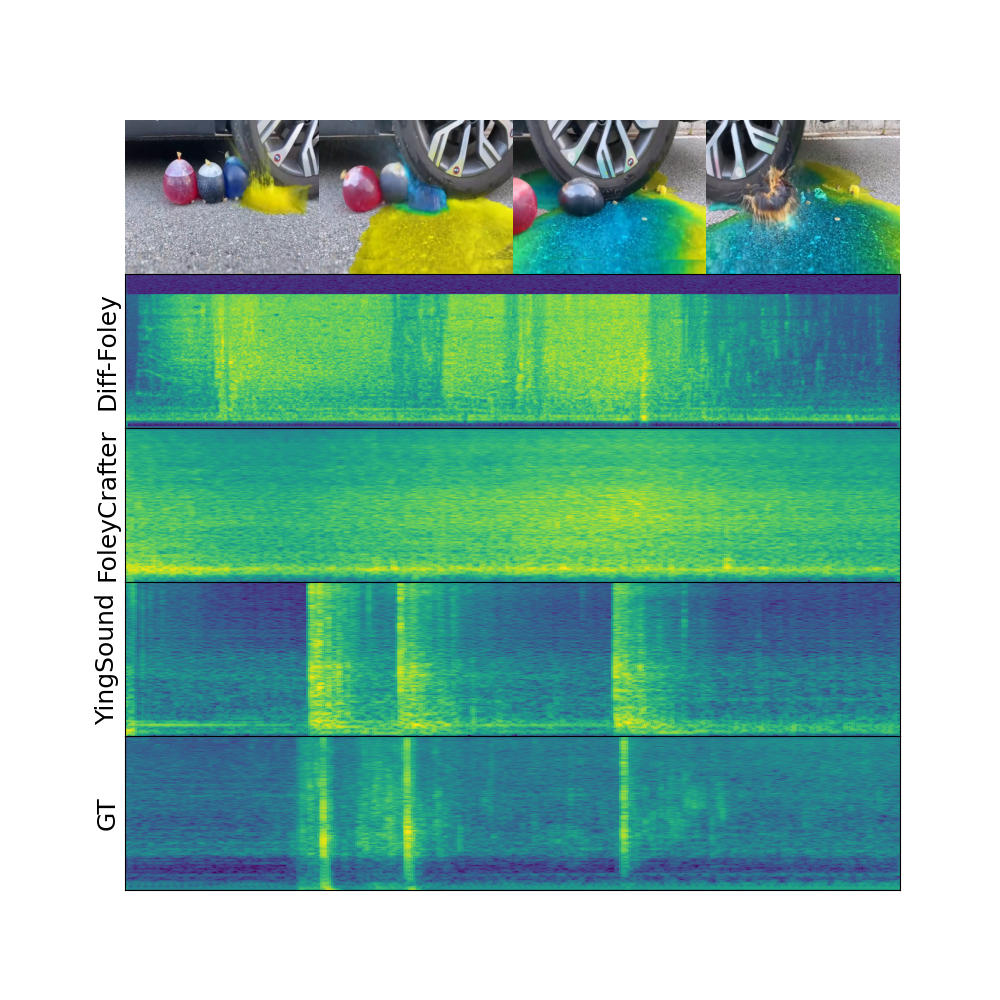}	
		}
		\subfigure{
			\includegraphics[width=0.49\linewidth,height=1.2in]{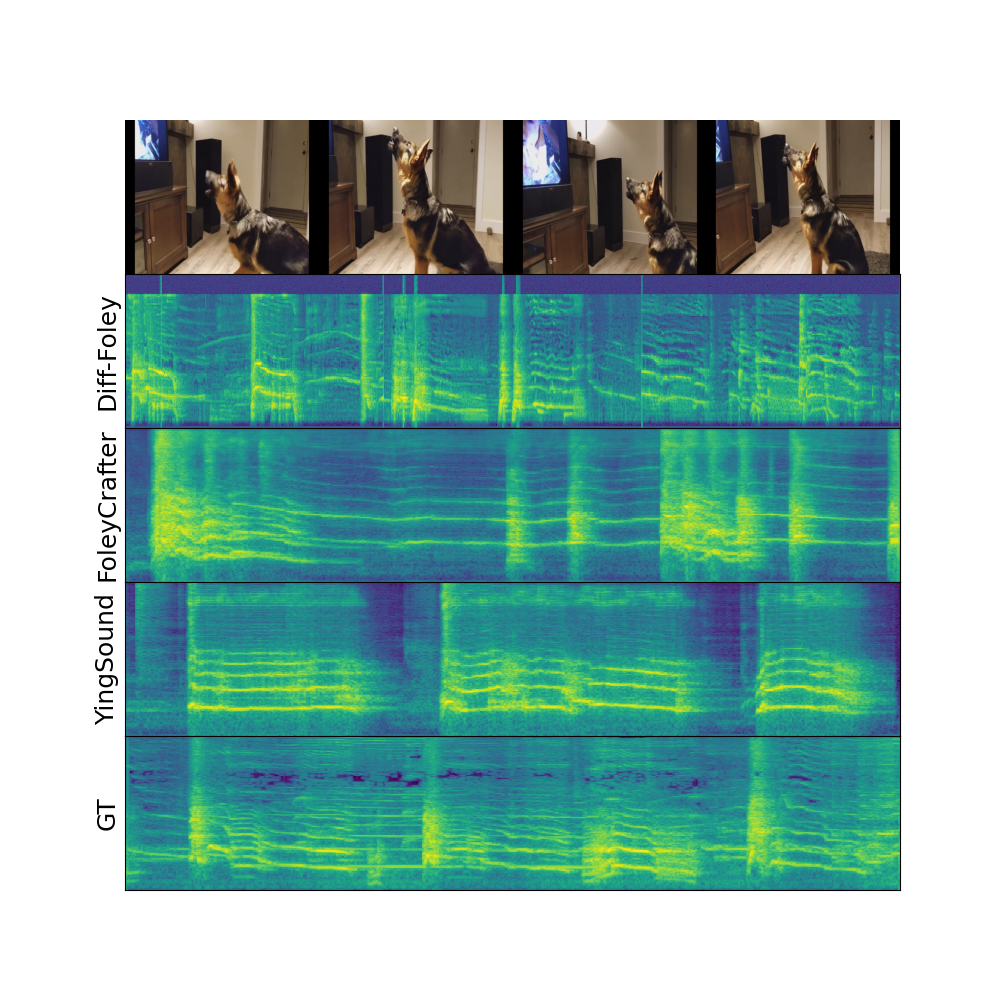}
		}
	\end{minipage}
	\vskip -0.3cm
	\begin{minipage}{1\linewidth}
		\subfigure{
			\includegraphics[width=0.49\linewidth,height=1.2in]{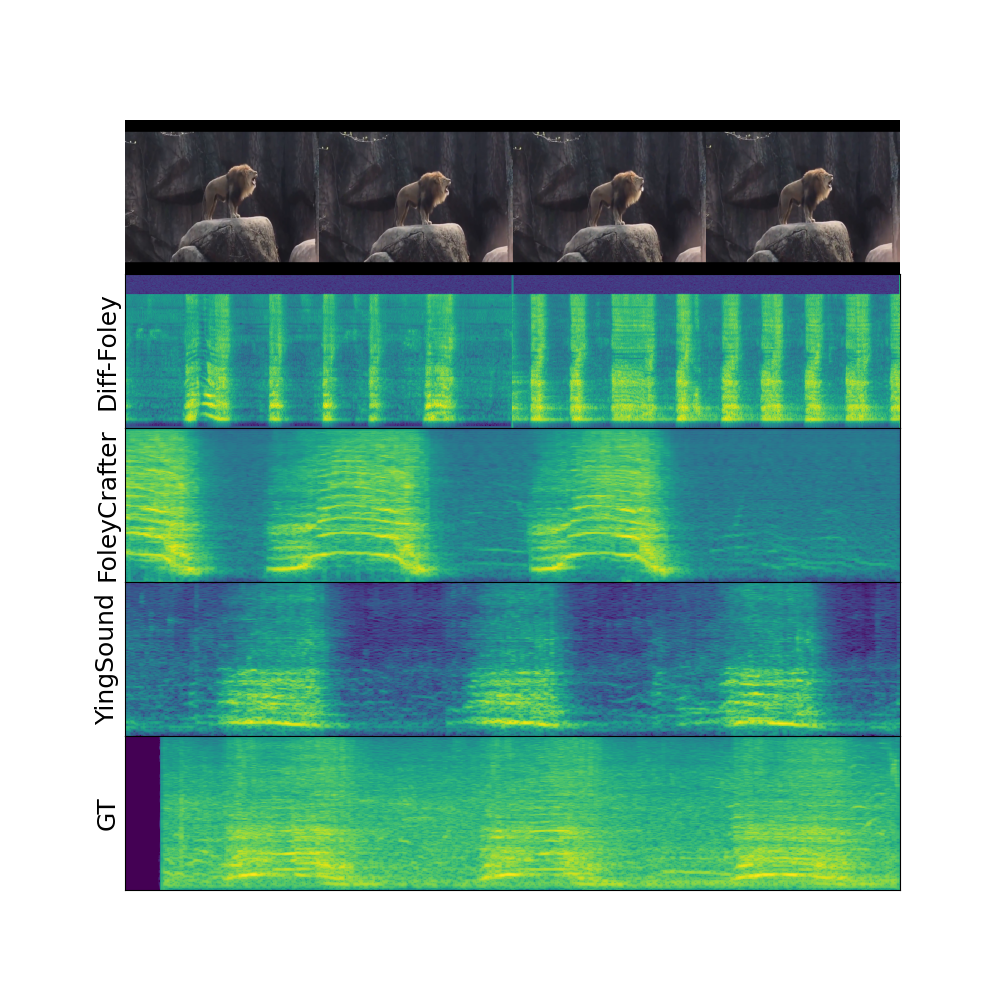}
		}
		\subfigure{
			\includegraphics[width=0.49\linewidth,height=1.2in]{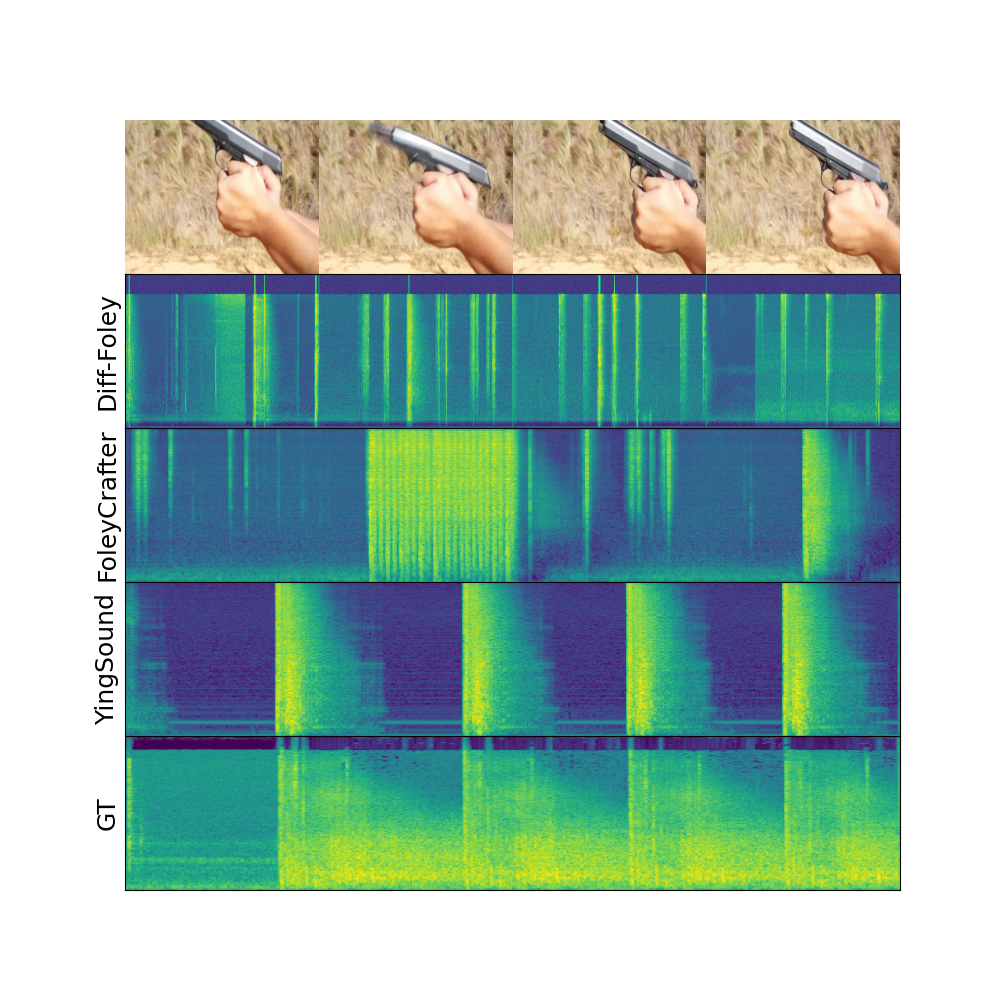}
		}
	\end{minipage}
	\vspace{-0.18in}
	\caption{\textbf{Temporal Alignment comparison.}}
	\vspace{-0.2in}
	\label{fig:1234}
\end{figure}

\subsection{Video-to-Audio Experiments} We present a quantitative comparison of semantic alignment and audio quality in the VGGSound-test dataset. The VGGSound-test dataset comprises 15,446 videos collected from YouTube, covering a diverse range of genres. The results show that the YingSound model, particularly when integrated with text input, exhibits superior performance across multiple objective metrics, including semantic alignment, temporal alignment with visual input, and better audio fidelity. 
\subsubsection{Audio Fidelity} 
As shown in Table \ref{tab:2}, the YingSound w. text model achieves the lowest FAD of 0.78, indicating a high level of audio realism, and the lowest FD of 6.28, suggesting excellent audio quality. It also achieves the lowest KL-sigmoid with a score of 3.97, which indicates the overall high-quality of sound effects generated based on video. Furthermore, the YingSound w. text model stands out with an IS of 14.02, which is a testament to the diversity and quality of the audio content it generates. 

\subsubsection{Semantic Alignment} 
It is important to note that when it comes to the CLIP, which measures semantic alignment, the YingSound w. text model achieves the highest score of 16.86, indicating a very strong correlation between the generated audio and the corresponding video. This highlights that both models, despite differences in their input, perform exceptionally well in ensuring that the generated audio is semantically aligned with the accompanying text. The YingSound w. text model's advantage in the CLIP suggests that it may have a slight edge in capturing the nuances of the text description for semantic alignment.  
\subsubsection{Temporal Alignment} 
The AV metric evaluates the synchronization between each segment of the input audio and its corresponding video segment. In Table \ref{tab:2}, the YingSound w. text, YingSound w/o text, and FoleyCrafter w/o text models all achieve an impressive AV score of 0.25, tying for the highest score among the compared models. This demonstrates that YingSound models are capable of generating audio with excellent temporal alignment to the video, highlighting their strong performance in this critical aspect of cross-modal generation quality.  \\
As shown in Figure \ref{fig:1234}, we compare ground truth (GT), Diff-Foley, FoleyCrafter, and our own YingSound models. It can be observed that the audio generated by YingSound exhibits nearly identical temporal alignment to the ground truth, demonstrating that our model achieves exceptional temporal alignment performance. However, Diff-Foley frequently generates either more or fewer sound effects than the ground truth. The results of the FoleyCrafter generation exhibit some misalignment in terms of timing.

\begin{figure}[htbp]
    \centering
    \includegraphics[width=0.9\textwidth]{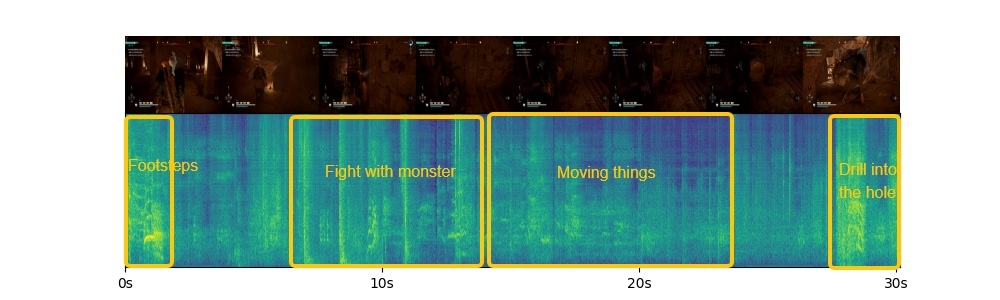}
    \vspace{0.5pt} 

    \includegraphics[width=0.9\textwidth]{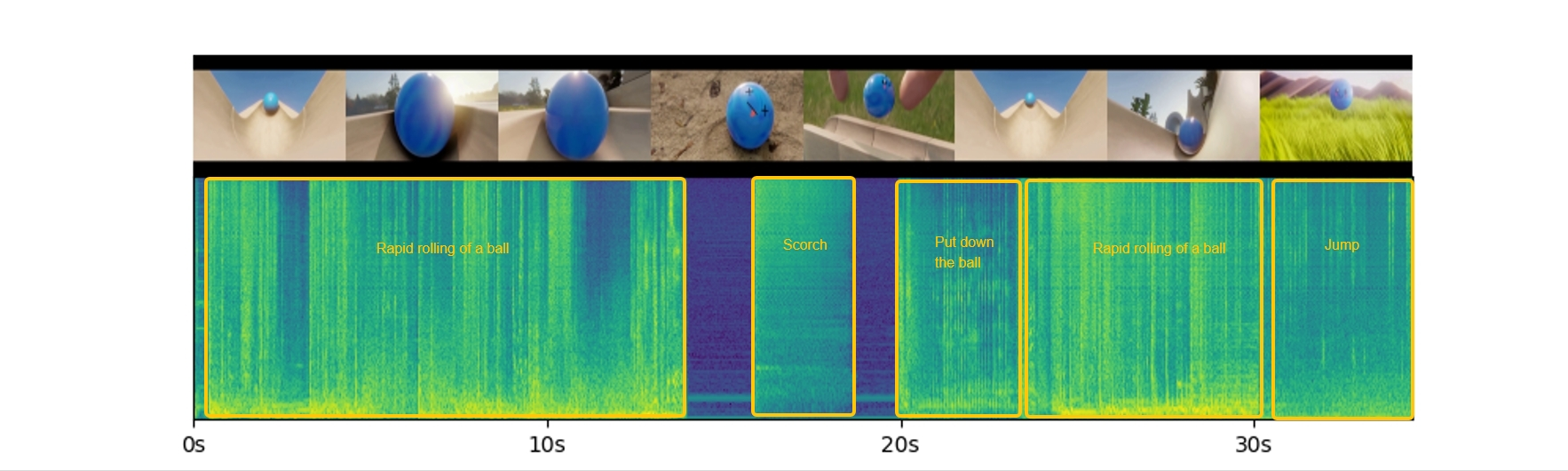}
    \vspace{0.5pt} 

    \includegraphics[width=0.9\textwidth]{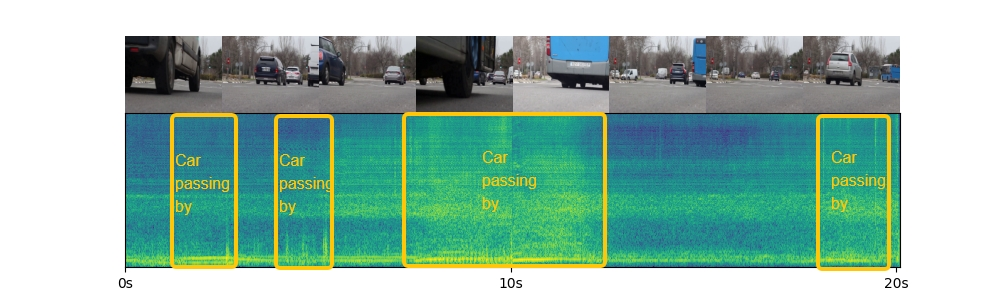}
    
    \caption{\textbf{Application visualization results of YingSound.} }
    \label{fig:5678}
\end{figure}

\section{Applications}
In few-shot settings, it is promising to apply industrial-level applications for long-duration scenes, including synchronized sound effects generation for the motions of game characters, animation videos, and short-term videos. Below are three application samples that represent the corresponding initial industrial-level applications. Examples of long-duration video inference and additional cases can be listened to on our Project Page.  

\subsection{Sound Effects for Game Character Motions}
In game-play, sound effects are essential for enhancing the realism and immersion of character movements, with the challenge lying in achieving precise synchronization between sound and actions at a finer level. Synchronizing audio with character actions provides players with a more dynamic and immersive gaming experience. The following samples showcase our demonstration of this promising application. We support the essential requirements for mid-duration sound effects generation for up to $30$ seconds with coarse time alignment to character motions (Figure \ref{fig:5678}). 

\subsection{Sound Effects for Animation Videos}
Sound effects are essential for enhancing generated videos by adding dynamic and contextually relevant audio that complements the visuals. The challenge lies in achieving semantic alignment between the generated audio and both the local and global dynamic visual content. Our model demonstrates high-quality sound effects generation for up to $30$ seconds for a single subject (Figure \ref{fig:5678}).    

\subsection{Sound Effects for Short Videos}
Sound effects are essential in enabling millions of short video creators and entrepreneurs to share their lives. A key factor is the adaptive production of sound effects that align with the varying rhythms of short videos, enhancing their impact and transmission potential towards zero-shot settings. YingSound demonstrates adaptive sound effects generation for up to $20$ seconds to some extent (Figure \ref{fig:5678}).

\section{Conclusion}
In this paper, we introduce YingSound for video-guided sound effects generation. We introduce a novel V2A model approach, which is the first version of a foundation model and features a learnable AVA structure. This model dynamically integrates high-resolution visual features with the corresponding audio features across multiple stages. To generate high-quality sound effects, we propose a multi-modal visual-audio CoT module that produces high-quality audio in few-shot settings. The various metrics on the benchmark, along with the temporal alignment charts, demonstrate our model's overall strong generation performance, high video semantic understanding, and exceptional temporal alignment capability. Additionally, we have built an industry-standard V2A dataset that encompasses a wide range of durations, including data from movies, games, and advertisements. \\

We are committed to continuously optimizing the existing V2A model, improving its performance across a wide range of application scenarios. Looking ahead, our team will focus on advancing its audio-related technical capabilities to provide an even more immersive and seamless user experience. One key area of development will be sound effects generation for game-play character motions, with a focus on sentimentally adaptive audio production including a variety of key elements: synchronization of action sequences, environment-specific adaptation, character-specific audio generation, layered audio design, and more.\\


\bibliographystyle{unsrt}
\bibliography{bib}
\end{document}